\documentclass[letterpaper, 10 pt, conference]{ieeeconf}
\IEEEoverridecommandlockouts

\usepackage[utf8]{inputenc}
\usepackage{amsmath}
\usepackage{mathrsfs}
\usepackage{amssymb}
\usepackage{color}
\usepackage{dsfont}
\usepackage{cite}
\usepackage{comment}
\usepackage{float}
\usepackage{graphicx}
\usepackage[font = footnotesize]{caption}
\usepackage[font = footnotesize]{subcaption}
\usepackage{bernsteinStyle}
\usepackage{float}
\usepackage{tikz}
\usepackage{booktabs}
\usepackage{tabularx}
\usepackage{multirow}
\usepackage{bm}
\usepackage{makecell}
\usepackage{indentfirst}
\newcolumntype{C}[1]{>{\centering\let\newline\\\arraybackslash\hspace{0pt}}m{#1}}

\usepackage{color, colortbl}
\definecolor{Yellow}{rgb}{1,1,0}

\usetikzlibrary{decorations.pathmorphing}
\usetikzlibrary{decorations.mypathmorphing}

\usetikzlibrary{shapes,arrows,calc,positioning}
\tikzstyle{bigblock} = [draw, fill=blue!20, rectangle, 
    minimum height=6em, minimum width=8em]
\tikzstyle{medblock} = [draw, fill=blue!20, rectangle, 
    minimum height=4em, minimum width=4em]    
\tikzstyle{mux} = [draw, fill=black!20, rectangle, 
    minimum height=5em, minimum width=0.1em]    
\tikzstyle{smallblock} = [draw, fill=blue!20, rectangle, 
    minimum height=3em, minimum width=4em]
\tikzstyle{sum} = [draw, fill=blue!20, circle, node distance=1cm]
\tikzstyle{signal} = [coordinate]
\tikzstyle{pinstyle} = [pin edge={to-,thin,black}]
\tikzstyle{block} = [draw, fill=blue!20, rectangle, 
    minimum height=3em, minimum width=6em]
\tikzstyle{blockS} = [draw, fill=blue!20, rectangle, 
    minimum height=3em, minimum width=4em]  
\tikzstyle{sum} = [draw, fill=blue!20, circle, node distance=1.5cm]
\tikzstyle{gain} = [draw, fill=blue!20, regular polygon, regular polygon sides = 3, node distance=1.25cm, shape border rotate = -90]
\tikzstyle{mult} = [draw, fill=blue!20, circle, node distance=1.25cm ,inner sep=0pt, minimum size = 0.3cm]
\tikzstyle{saturation block} = [draw,fill=blue!20, 
    	path picture={
    		\pgfpointdiff{\pgfpointanchor{path picture bounding box}{north east}}%
    		{\pgfpointanchor{path picture bounding box}{south west}}
    		\pgfgetlastxy\x\y
    		\tikzset{x=\x*.4, y=\y*.4}
    		\draw (-1,0) -- (1,0) (0,-1) -- (0,1); 
    		\draw (-1,-.7) -- (-.5,-.7) -- (.5,.7) -- (1,.7);
    	}]
\tikzstyle{sat atan} = [draw,fill=blue!20, 
        path picture={
          \pgfpointdiff{\pgfpointanchor{path picture bounding box}{north east}}%
            {\pgfpointanchor{path picture bounding box}{south west}}
          \pgfgetlastxy\x\y
          \tikzset{x=\x*.05, y=\y*0.3}
          \draw (-7.5,0) -- (7.5,0) (0,-1.2) -- (0,1.2); 
          \draw[scale = 1,domain=-7.5:7.5,smooth,variable=\x,black] plot ({\x},{(tanh((0.6*\x)))});
        }]
\tikzstyle{input} = [coordinate]
\tikzstyle{output} = [coordinate]

\tikzset{
  Speaker/.pic={
    \filldraw[fill=gray!40,pic actions] 
    (-15pt,0) -- 
      coordinate[midway] (-front) 
    (15pt,0) -- 
    ++([shift={(-6pt,8pt)}]0pt,0pt) coordinate (aux1) -- 
    ++(-18pt,0) coordinate (aux2) 
    -- cycle 
    (aux1) -- ++(0,6pt) -- coordinate[midway] (-back) ++(-18pt,0) -- (aux2);
  },
  Microphone/.pic={
    \filldraw[color=black, fill=gray!50](-2pt,0pt) rectangle (2pt, 4pt);
    \filldraw[color=black, fill=gray!50](-2pt,4pt) rectangle (2pt, 20pt);
  }
 }

\newcounter{example}

\title{\LARGE Experimental Application of Predictive Cost Adaptive Control\\ to Thermoacoustic Oscillations in a Rijke Tube}

\author{\large Juan A. Paredes and Dennis S. Bernstein
\thanks{Juan A. Paredes and Dennis S. Bernstein are with the Department of Aerospace Engineering, University of Michigan, Ann Arbor, MI, USA. {\tt\small \{jparedes, dsbaero\}@umich.edu}}
}

\begin{document}

\maketitle

\begin{abstract}
Model predictive control (MPC) has been used successfully in diverse applications.
As its name suggests, MPC requires a model for predictive optimization.
The present paper focuses on the application of MPC to a Rijke tube, in which a heating source and acoustic dynamics interact to produce self-excited oscillations.
Since the dynamics of a Rijke tube are difficult to model to a high level of accuracy,
the implementation of MPC requires leveraging data from the physical setup as well as knowledge about thermoacoustics, which is labor intensive and requires domain expertise.
With this motivation, the present paper uses predictive cost adaptive control (PCAC) for sampled-data control of an experimental Rijke-tube setup.
PCAC performs online closed-loop linear model identification for receding-horizon optimization based on the backward propagating Riccati equation.
In place of analytical modeling, open-loop experiments are used to create a simple emulation model, which is used for choosing PCAC hyperparameters.
PCAC is applied to the Rijke-tube setup under various experimental scenarios.
\end{abstract}

%

\section{Introduction}

In many ways, a Rijke tube provides an ideal laboratory testbed for developing and assessing control techniques.
A Rijke tube consists of a glass tube and a heating element, which induces thermoacoustic oscillations due to the interaction between the nonlinear heat release and the linear acoustic dynamics \cite{rijke1859,rayleigh1878,heckl1990,bittanti2002,epperlein2015,rubio2015,deandrade2020}.
These dynamics are self-excited, which means that a constant voltage to the heating element leads to pressure oscillations in the tube.
For feedback control, a microphone and speaker are needed, as well as a processor and amplifiers to implement control algorithms.
The resulting setup is inexpensive to build and operate, and is immune to damage in the event of instability.

The dynamics of a Rijke tube, which involve both heat transfer and acoustics, are sufficiently complex that they are not easily modeled to a sufficiently high degree of accuracy to facilitate model-based control.
Nevertheless, linear control methods based on linear system identification and linearized analytical models have been applied extensively to this application despite the fact that the system is inherently nonlinear \cite{heckl1988, annaswamy2000,illingworth2010,epperlein2015,zalluhoglu2016, deandrade2017,juanACCrijke,RijkeTCST}.
This mismatch raises fundamental questions about the serendipitous effectiveness of linear methods on this nonlinear system.

The present paper further explores the effectiveness of linear modeling/linear controllers for the Rijke tube by applying model predictive control (MPC).
MPC has been used successfully in diverse applications \cite{lee2011,darby2012,afram2014,incremona2017,torrente2021,aliramezani2022}, and is likely the most effective modern control technique.
As its name suggests, however, MPC requires a model for predictive optimization.
Implementations of MPC to Rijke-tube experiments require leveraging knowledge about the specific Rijke-tube configuration and thermoacoustics.
For example, \cite{jarmolowitz2012conf,jarmolowitz2012journal} require knowledge of the Rijke-tube configuration to develop a robust observer, while \cite{shariati2014} requires knowledge of the Helmholtz mode corresponding to the open-loop, Rijke-tube self-oscillations.
Since the dynamics of a Rijke tube are difficult to model to a high level of accuracy, this testbed provides a challenging application for MPC.

With this motivation, the present paper uses predictive cost adaptive control (PCAC) for sampled-data control of an experimental Rijke-tube setup.
PCAC performs online closed-loop linear system identification identification;  the identified model is then used as the basis for receding-horizon optimization.
The system identification is performed during closed-loop operation by recursive least squares (RLS) \cite{islam2019recursive,mohseni2022recursive}. 
For receding-horizon optimization, quadratic programming (QP) is used in \cite{islamPCAC}.
Since state and control constraints are not crucial for the Rijke-tube experiments, the present paper uses the backward propagating Riccati equation (BPRE) \cite{kwon2006receding,kwonpearson} in place of QP.
An additional advantage of BPRE over QP is computational simplicity, which provides the means to implement PCAC at a high sample rate (1 kHz).

The Rijke-tube setup is used in \cite{juanACCrijke,RijkeTCST} as a testbed for retrospective cost adaptive control (RCAC).
One of the goals of the present paper is to assess the performance of PCAC relative to RCAC.
As in \cite{juanACCrijke,RijkeTCST}, in place of analytical modeling, open-loop experiments are used to create a simple emulation model, which is used for choosing PCAC hyperparameters.
PCAC is applied to the Rijke-tube setup under various experimental scenarios.

The contents of the paper are as follows.
Section \ref{sec:prob} provides a statement of the control problem, which involves continuous-time dynamics under sampled-data feedback control.
The control input is subjected to magnitude saturation.
Section \ref{sec:PCAC} describes the predictive control law considered in this paper for suppression.
Section \ref{sec:Rijke_Exp} presents the Rijke-tube setup.
Section \ref{sec:Rijke_Exp_Res} presents physical closed-loop results using the Rijke-tube setup.
Finally, Section \ref{sec:conclusion} presents conclusions. 

{\bf Notation:}
$x_{(i)}$ denotes the $i$th component of $x\in\BBR^n.$
${\rm sprad}(A)$ denotes the spectral radius of $A\in\BBR^{n\times n}.$
The symmetric matrix $P\in\BBR^{n\times n}$ is positive semidefinite (resp., positive definite) if all of its eigenvalues are nonnegative (resp., positive).
$\vek X\in\BBR^{nm}$ denotes the vector formed by stacking the columns of $X\in\BBR^{n\times m}$, and 
$\otimes$ denotes the Kronecker product.
$I_n$ is the $n \times n$ identity matrix, $0_{n\times m}$ is the $n\times m$ zeros matrix, and $\mathds{1}_{n\times m}$ is the $n\times m$ ones matrix.


\section{Statement of the Control Problem}\label{sec:prob}

\begin{figure} [h!]
    \centering
    \vspace{0.25em}
    \resizebox{0.8\columnwidth}{!}{%
    \begin{tikzpicture}[>={stealth'}, line width = 0.25mm]

    \node [input, name=ref]{};
    \node [smallblock, rounded corners, right = 0.5cm of ref , minimum height = 0.6cm, minimum width = 0.7cm] (controller) {$G_{\rmc,k}$};
    \node[smallblock, rounded corners, right = 1cm of controller, minimum height = 0.6cm, minimum width = 0.5cm](sat_Blk){$\sigma$};
    \node [smallblock, rounded corners, right = 0.75cm of sat_Blk, minimum height = 0.6cm , minimum width = 0.5cm] (DA) {ZOH};
    
    \node [smallblock, fill=green!20, rounded corners, right = 0.75cm of DA, minimum height = 0.6cm , minimum width = 1cm] (system) {$\SM$};
    \node [output, right = 0.5cm of system] (output) {};
    \node [input, below = 0.9cm of system] (midpoint) {};
    
    \draw [->] (controller) -- node [above] {$u_{{\rm req},k}$} (sat_Blk);\
    \draw [->] (sat_Blk) -- node [above] {$u_k$} (DA);\
    \draw [->] (DA) -- node [above] {$u (t)$} (system);
    
    \node[circle,draw=black, fill=white, inner sep=0pt,minimum size=3pt] (rc11) at ([xshift=-2.5cm]midpoint) {};
    \node[circle,draw=black, fill=white, inner sep=0pt,minimum size=3pt] (rc21) at ([xshift=-2.8cm]midpoint) {};
    \draw [-] (rc21.north east) --node[below,yshift=.55cm]{$T_\rms$} ([xshift=.3cm,yshift=.15cm]rc21.north east) {};
    
    \draw [->] (system) -- node [name=y, near end]{} node [very near end, above] {$y (t)$}(output);
    
    \draw [-] (y.west) |- (midpoint);
    \draw [-] (midpoint) -| node [very near end, above, xshift=-0.7cm] {$y_k$} (rc11.east);
    \draw [->] (rc21) -| ([xshift = -0.5cm]controller.west) -- (controller.west);
    
    \end{tikzpicture}
    }  
    \caption{Sampled-data implementation of predictive controller for stabilization of continuous-time system $\SM.$
    All sample-and-hold operations are synchronous.
    The predictive controller $G_{\rmc,k}$ generates the requested discrete-time control $u_{{\rm req},k}\in\BBR^m$ at each step $k$.
    The implemented discrete-time control is $u_k=\sigma(u_{{\rm req},k})$, where $\sigma\colon\BBR^m\to\BBR^m$ represents control-magnitude saturation.
    The resulting continuous-time control $u(t)$ is generated by applying a zero-order-hold operation to $u_k$.
    For this work, $\SM$ represents 
    the Rijke-tube setup introduced in Section \ref{sec:Rijke_Exp} for physical experiments.}
    \label{fig:PC_CT_blk_diag}
\end{figure}
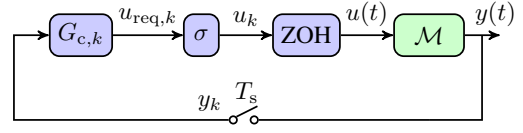

To reflect the practical implementation of digital controllers for physical systems, we consider continuous-time dynamics under sampled-data control using discrete-time predictive controllers.
In particular, we consider the control architecture shown in Figure \ref{fig:PC_CT_blk_diag}, where $\SM$ is the target continuous-time system, $t\ge 0$, $u(t)\in\BBR^m$ is the control, and $y(t)\in\BBR^p$ is the output of $\SM,$ which is sampled to produce the measurement $y_k \in \BBR^p,$ which, for all $k\ge0,$ is given by
\begin{equation}
    y_k \isdef  y(k T_\rms),
\end{equation}
where  $T_\rms>0$ is the sample time.
The predictive controller, which is updated at each step $k,$ is denoted by $G_{\rmc,k}$.
The input to $G_{\rmc,k}$ is $y_k$, and its output at each step $k$ is the requested discrete-time control $u_{{\rm req},k}\in\BBR^m.$
Since the response of a real actuator is subjected to hardware constraints, the implemented discrete-time control is 
\begin{equation}
 u_k\isdef \sigma(u_{{\rm req},k}),  
 \label{eq:ukksat}
\end{equation}
where $\sigma\colon\BBR^m\to\BBR^m$ is the control-magnitude saturation function  
\begin{equation}
    \sigma(u) \isdef  \matl \bar{\sigma}(u_{(1)}) \\ \vdots \\ \bar{\sigma} (u_{(m)})\matr,
    \label{eq:Hsat1}
\end{equation}
where $\bar\sigma\colon \BBR\to\BBR$ is defined by 
\begin{equation}
    \bar\sigma(u)\isdef \begin{cases} u_{\max},&u> u_{\max},\\
    u,& u_{\min}\le u \le u_{\max},\\
    u_{\min}, & u<u_{\min},\end{cases}
\end{equation}
and $u_{\min},u_{\max}\in\BBR$ are the lower and upper magnitude saturation levels, respectively.
Rate (move-size) saturation can also be considered.
The continuous-time control signal $u(t)$ applied to the structure is generated by applying a zero-order-hold operation to $u_k,$ that is,
for all $k\ge0,$ and, for all $t\in[kT_\rms, (k+1) T_\rms),$ 
\begin{equation}
    u(t) = u_k.
\end{equation}
The objective of the predictive controller is to yield an input signal that minimizes the output of the continuous-time system, that is, yield $u(t)$ such that $\lim_{t \to \infty} y(t) = 0.$

For this work, rate saturation is not considered, and $\SM$ represents
the experimental Rijke-tube setup introduced in Section \ref{sec:Rijke_Exp} for physical experiments.

\section{Predictive Cost Adaptive Control} \label{sec:PCAC}

The PCAC algorithm is presented in this section.
Subsection \ref{subsec:ID} describes the technique used for online identification, namely, RLS with variable-rate forgetting based on the F-test \cite{mohseni2022recursive}.
Subsection \ref{subsec:bocf} presents the block observable canonical form (BOCF), which is used to represent the input-output dynamics model as a state space model whose state is given explicitly in terms of inputs, outputs, and model-coefficient estimates.
Subsection \ref{subsec:bpre} reviews the BPRE technique for receding-horizon optimization.
Using BOCF, the full-state feedback controller obtained using BPRE is implementable as an output-feedback dynamic compensator.

\subsection{Online Identification Using Recursive Least Squares with Variable-Rate Forgetting
Based on the F-Test} \label{subsec:ID}

Let $\hat n\ge 0$ and, for all $k\ge 0,$ let $F_{\rmm,1,k},\hdots, F_{\rmm,\hat n,k}\in\BBR^{p\times p}$ and $G_{\rmm,1,k},\hdots, G_{\rmm,\hat n,k}\in\BBR^{p\times m}$ be the coefficient matrices to be estimated using RLS.
Furthermore, let $\hat y_k\in\BBR^p$ be an estimate of $y_k$ defined  by
\begin{equation}
\hat y_k\isdef -\sum_{i=1}^{\hat n}  F_{\rmm,i,k}   y_{k-i} + \sum_{i=1}^{\hat n} {G}_{\rmm,i,k} u_{k-i},
\label{eq:yhat}
\end{equation}
where  
\begin{gather}
   y_{-\hat n}=\cdots= y_{-1}=0,\\ u_{-\hat n}=\cdots=u_{-1}=0. 
\end{gather}
Using the identity ${\rm vec} (XY) = (Y^\rmT \otimes I) {\rm vec} X,$ it follows from  \eqref{eq:yhat} that, for all $k\ge 0,$ 
\begin{equation}
    \hat y_k = \phi_k \theta_k,
    \label{eq:yhat_phi}
\end{equation}
where
\begin{align}
     %
     \theta_k \isdef & \ \matl \theta_{F_\rmm, k}^\rmT & \theta_{G_\rmm, k}^\rmT \matr^\rmT \in\BBR^{\hat np(m+p)},\\
     \theta_{F_\rmm, k} \isdef & \ {\rm vec}\matl  F_{\rmm,1,k}&\cdots& F_{\rmm,\hat n,k} \matr \in\BBR^{\hat n p^2}, \\
     \theta_{G_\rmm, k} \isdef & \ {\rm vec}\matl  G_{\rmm,1,k}&\cdots& G_{\rmm,\hat n,k} \matr \in\BBR^{\hat n pm}, \\
     \phi_k \isdef & \matl -  y_{k-1}^\rmT&\cdots&- y_{k-\hat n}^\rmT&u_{k-1}^\rmT&\cdots& u_{k-\hat n}^\rmT\matr\otimes I_p \nn \\ &\in\BBR^{p\times \hat np(m+p)}.
\label{eq:phi_kkka}
 \end{align}

To determine the update equations for $\theta_k$, for all $k\ge 0$, define $e_k\colon\BBR^{\hat np(m+p)}\to\BBR^p$ by
\begin{equation}
    e_k(\bar \theta) \isdef y_k - \phi_k \bar \theta,
    \label{eq:ekkea}
\end{equation}
where $\bar \theta\in\BBR^{\hat np(m+p)}.$ 
Using  \eqref{eq:yhat_phi}, the \textit{identification error} at step $k$ is defined by
\begin{equation}
 e_k(\theta_k)= y_k-\hat y_k.  
\end{equation}
For all $k\ge 0$, the RLS cumulative cost $J_k\colon\BBR^{\hat np(m+p)}\to[0,\infty)$ is defined by \cite{islam2019recursive}
\begin{equation}
J_k(\bar \theta) \isdef \sum_{i=0}^k \frac{\rho_i}{\rho_k} e_i^\rmT(\bar \theta) e_i(\bar \theta) + \frac{1}{\rho_k} (\bar\theta -\theta_0)^\rmT \Psi_0^{-1}(\bar\theta-\theta_0),
\label{Jkdefn}
\end{equation}
where $\Psi_0\in\BBR^{\hat np(m+p)\times \hat np(m+p)}$ is  positive definite, $\theta_0\in\BBR^{\hat n p(m+p)}$ is the initial estimate of the coefficient vector, and, for all $i\ge 0,$
\begin{equation}
  \rho_i \isdef \prod_{j=0}^i \lambda_j^{-1}.  
\end{equation}
For all $j\ge 0$, the parameter $\lambda_j\in(0,1]$ is the forgetting factor defined by $\lambda_j\isdef\beta_j^{-1}$, where
\begin{equation}
 \beta_j \isdef \begin{cases}
     1, & j<\tau_\rmd,\\
     1 + \eta \bar{\beta}_j,& j\ge \tau_\rmd,
 \end{cases}    
\end{equation}
\footnotesize
\begin{equation}
    \bar{\beta}_j \isdef g(e_{j-\tau_\rmd}(\theta_{j-\tau_\rmd}),\hdots,e_j(\theta_j)) \textbf{1}\big(g(e_{j-\tau_\rmd}(\theta_{j-\tau_\rmd}),\hdots,e_j(\theta_j))\big),
\end{equation}
\normalsize
and  $\tau_\rmd> p$, $\eta>0$,  $\textbf{1}\colon \BBR\to\{0,1\}$ is the unit step function, and $g$ is a function of past RLS identification errors.
To determine $g$ when $p=1$, let $\tau_\rmn\in[p,\tau_\rmd)$, and let  $\sigma_{k,\tau_\rmd}^2$ and $\sigma_{k,\tau_\rmn}^2$ be the variances of past RLS prediction-error sequences $\{e_{k-\tau_\rmd}(\theta_{k-\tau_\rmd}),\hdots,e_{k}( \theta_{k})\}$ and $\{e_{k-\tau_\rmn}( \theta_{k-\tau_\rmn}),\hdots,e_{k}( \theta_{k})\}$, respectively.
In this case, $g\colon \BBR^p\times\cdots\times \BBR^p$ is defined by
\begin{equation}
    g(e_{k-\tau_\rmd}(\theta_{k-\tau_\rmd}),\hdots,e_k(\theta_k))\isdef \sqrt{\frac{\sigma_{k,\tau_\rmn}^2}{\sigma_{k,\tau_\rmd}^2}} - \sqrt{F_{\tau_\rmn,\tau_\rmd}^{\rm inv}(1-\alpha)},
    \label{eq:gg}
\end{equation}
where $\alpha\in (0,1]$ is the \textit{significance level}, and  $F^{\rm inv}_{\tau_\rmn,\tau_\rmd}(x)$ is the inverse cumulative distribution function of the F-distribution with degrees of freedom $\tau_\rmn$ and $\tau_\rmd.$
Note \eqref{eq:gg} enables forgetting when $\sigma_{\tau_\rmn}^2$ is statistically larger than $\sigma_{\tau_\rmd}^2.$
Moreover, larger values of the significance level $\alpha$ cause the level of forgetting to be more sensitive to changes in the ratio of $\sigma_{\tau_\rmn}^2$ to $\sigma_{\tau_\rmd}^2$.

When $p> 1,$ instead of variances $\sigma_{k,\tau_\rmd}$ and $\sigma_{k,\tau_\rmn}$, we consider covariance matrices $\Sigma_{k,\tau_\rmd}$ and $\Sigma_{k,\tau_\rmn}$, and thus the product  $\Sigma_{k,\tau_\rmn}\Sigma_{k,\tau_\rmd}^{-1}$ replaces the ratio $\sigma_{k,\tau_\rmn}^2/\sigma_{k,\tau_\rmd}^2$. 
In this case, $g\colon \BBR^{p}\times \cdots\times \BBR^{p} $ is defined by 
\begin{align}
    g(e_{k-\tau_\rmd}&(\theta_{k-\tau_\rmd}),\hdots,e_k(\theta_k)) \nn \\
    &\isdef \sqrt{\frac{\tau_\rmn}{c\tau_\rmd}\tr \big(\Sigma_{k,\tau_\rmn}\Sigma_{k,\tau_\rmd}^{-1}\big)} - \sqrt{F^{\rm inv}_{p\tau_\rmn,b}(1-\alpha)},
\end{align}
where 
\begin{gather}
    a\isdef \frac{(\tau_\rmn + \tau_\rmd - p -1)(\tau_\rmd-1)}{(\tau_\rmd-p-3)(\tau_\rmd - p)},\\
    b\isdef 4 + \frac{p\tau_\rmn +2 }{a-1},\quad c\isdef \frac{p\tau_\rmn(b-2)}{b(\tau_\rmd-p-1)}.
\end{gather}

Finally, for all $k\ge0$, the unique global minimizer of $J_k$ is given by \cite{islam2019recursive}
\begin{equation}
    \theta_{k+1} = \theta_k +\Psi_{k+1} \phi_k^\rmT (y_k - \phi_k \theta_k),
\end{equation}
where 
\begin{align}
  \Psi_{k+1} &\isdef  \beta_k \Psi_k - \beta_k \Psi_k \phi_k^\rmT (\tfrac{1}{\beta_k}I_p + \phi_k  \Psi_k \phi_k^\rmT)^{-1} \phi_k  \Psi_k,
\end{align}
and $\Psi_0$ is the performance-regularization weighting in \eqref{Jkdefn}.
Additional details concerning RLS with forgetting based on the F-distribution are given in \cite{mohseni2022recursive}.


\subsection{Input-Output Model and the Block Observable Canonical Form} \label{subsec:bocf}

Considering the estimate $\hat y_k$ of $y_k$ given by \eqref{eq:yhat}, it follows that, for all $k\ge0,$
\begin{equation}
y_{k} \approx -\sum_{i=1}^{\hat n}  F_{\rmm,i,k}   y_{k-i} + \sum_{i=1}^{\hat n} {G}_{\rmm,i,k} u_{k-i}.
\label{eq:ykapp}
\end{equation}
Viewing \eqref{eq:ykapp} as an equality, it follows that, for all $k\ge 0,$ the BOCF state-space realization of \eqref{eq:ykapp} is given by  \cite{polderman1989state}
\begin{align}
 x_{\rmm,k+1} &=   A_{\rmm,k}  x_{\rmm,k} +  B_{\rmm,k} u_k,\label{eq:xmssAB}\\
 y_{k} &=  C_\rmm   x_{\rmm, k},
\label{eq:yhatCxm}
\end{align}
where
\begin{gather}
 A_{\rmm,k} \isdef \matl - F_{\rmm,1,k+1} & I_p & \cdots & \cdots & 0_{p\times p}\\
- F_{\rmm,2,k+1} & 0_{p\times p} & \ddots & & \vdots\\
\vdots & {\vdots} & \ddots & \ddots & 0_{p\times p} \\
\vdots & \vdots &  & \ddots & I_p\\
- F_{\rmm,\hat n,k+1} & 0_{p\times p} & \cdots &\cdots & 0_{p\times p}
\matr\in\BBR^{\hat np\times \hat n p},\label{eq:ABBA_1}\\
B_{\rmm,k}\isdef \matl  G_{\rmm,1,k+1} \\
 G_{\rmm,2,k+1}\\
\vdots\\
 G_{\rmm,\hat n,k+1}
\matr \in\BBR^{\hat n p \times m},\label{eq:ABBA_2}
\end{gather}
\begin{gather}
  C_\rmm\isdef \matl I_p & 0_{p\times p} & \cdots & 0_{p\times p} \matr\in\BBR^{p\times \hat n p},
  \label{eq:cmmx}
\end{gather}
and 
\begin{equation}
    x_{\rmm,k} \isdef \matl x_{\rmm,k(1)}\\\vdots\\ x_{\rmm,k(\hat n)}\matr \in\BBR^{\hat n p},
    \label{eq:xmkkx}
\end{equation}
where
\begin{align}
 x_{\rmm,k(1)} \isdef  y_{k},
\label{eq:x1kk}
\end{align}
and, for all $j=2,\ldots,\hat n,$
\begin{align}
 x_{\rmm,k(j)} \isdef & -\sum_{i=1}^{\hat n -j +1}  F_{\rmm,i+j-1,k+1}  y_{k-i} \nn \\
 &+ \sum_{i=1}^{\hat n -j+1}  G_{\rmm,i+j-1,k+1} u_{k-i}.
 \label{eq:xnkk}
\end{align}
Note that multiplying both sides of \eqref{eq:xmssAB} by $C_\rmm$ and using \eqref{eq:yhatCxm}--\eqref{eq:xnkk} implies that, for all $k\ge 0,$
\begin{align}
     y_{k+1}=&C_\rmm x_{\rmm,k+1}\nn\\
     =& C_\rmm(A_{\rmm,k}  x_{\rmm,k} +  B_{\rmm,k} u_k)\nn\\
    =&-F_{\rmm,1,k+1}  x_{\rmm,k(1)} + x_{\rmm,k(2)} +G_{\rmm,1,k+1} u_k \nn\\
    =&-F_{\rmm,1,k+1}  y_{k}  -\sum_{i=1}^{\hat n -1}  F_{\rmm,i+1,k+1}  y_{k-i} \nn\\
    &+ \sum_{i=1}^{\hat n -1}  G_{\rmm,i+1,k+1} u_{k-i}+G_{\rmm,1,k+1} u_k \nn\\
    =&  -\sum_{i=1}^{\hat n }  F_{\rmm,i,k+1}  y_{k+1-i} + \sum_{i=1}^{\hat n }  G_{\rmm,i,k+1} u_{k+1-i},
\end{align}
which is approximately equivalent to \eqref{eq:ykapp} with $k$ in \eqref{eq:ykapp} replaced by $k+1$.


\subsection{Receding-Horizon Control with Backward-Propagating Riccati Equation (BPRE)} \label{subsec:bpre}

In this section, we use receding-horizon optimization to determine the requested control $u_{{\rm req},k+1}$ and thus the implemented control $u_{k+1}$, as discussed in Section \ref{sec:prob}.
Let $\ell\ge 1$ be the horizon, and, for all $k\ge0$ and all $j=1,\ldots,\ell,$ consider the state-space prediction model
\begin{equation}
    x_{\rmm,k|j+1} =   A_{\rmm,k}  x_{\rmm,k|j} +  B_{\rmm,k}  u_{{\rm req},k|j},
\end{equation}
where $A_{\rmm,k}$ and $B_{\rmm,k}$ are given by \eqref{eq:ABBA_1} and \eqref{eq:ABBA_2}, respectively,
$x_{\rmm,k|j}\in\BBR^{\hat n p}$ is the $j$-step predicted state, $u_{{\rm req},k|j}\in\BBR^m$ is the $j$-step predicted control, and the initial conditions are
\begin{equation}
 x_{\rmm,k|1} \isdef  x_{\rmm,k+1},\quad u_{{\rm req},k|1}\isdef  u_{{\rm req},k+1}. \label{eq:xmtuk1} 
\end{equation}
Note that, at each step  $k\ge0$, after obtaining the measurement $y_k$, $x_{\rmm,k+1}$ is computed using \eqref{eq:xmssAB}, where $u_k$ is the implemented control at step $k$ given by \eqref{eq:ukksat}.
Furthermore,  $u_{{\rm req},k+1}$,
which is determined below, is the requested discrete-time control at step $k+1$.
For all $k\ge0$, define the performance index
\begin{align}
\SJ_k(&u_{{\rm req},k|1},\hdots, u_{{\rm req},k|\ell}) \nn \\ 
\isdef & \ \half \sum_{j=1}^{\ell} ( x_{\rmm,k|j}^\rmT R_{1,k|j}  x_{\rmm,k|j} +  u_{{\rm req},k|j}^\rmT R_{2,k|j}   u_{{\rm req},k|j}) \nn \\
&+ \half  x_{\rmm,k|\ell+1}^\rmT P_{k|\ell+1} x_{\rmm,k|\ell+1},\label{SJk}
\end{align}
where the terminal weighting $P_{k|\ell+1}\in\BBR^{\hat np\times \hat np}$ is positive semidefinite and,
for all  $j=1,\ldots,\ell,$  
$R_{1,k|j}\in\BBR^{\hat np\times \hat n p}$ is the positive semidefinite state weighting and $R_{2,k|j}\in\BBR^{m\times m}$ is the positive definite control weighting.
The first term in \eqref{SJk} can be written as
\begin{equation}
    x_{\rmm,k|j}^\rmT R_{1,k|j}  x_{\rmm,k|j} = z_k^\rmT z_k,
\end{equation}
where $z_k\in\BBR^p$ is defined by
\begin{equation}
    z_k \isdef E_{1,k|j}x_{\rmm,k|j},
\end{equation}
and $E_{1,k|j}\in\BBR^{p\times \hat n p}$ is defined such that
\begin{equation}
    R_{1,k|j} = E_{1,k|j}^\rmT E_{1,k|j}.
\end{equation}
With this notation, $z_k$ is the performance variable.

For all $k\ge0$ and all $j=\ell,\ell-1,\ldots,2,$ let $P_{k|j}$ be given by
%
%
\begin{align}
P_{k|j} =& \ A_{\rmm,k}^\rmT P_{k|j+1}   \left(A_{\rmm,k} -  B_{\rmm,k} \Gamma_{k|j} \right) + R_{1,k|j}, \\
\Gamma_{k|j} \isdef & \ (R_{2,k|j} +   B_{\rmm,k}^\rmT P_{k|j+1}  B_{\rmm,k})^{-1}   B_{\rmm,k}^\rmT P_{k|j+1}   A_{\rmm,k}.
\end{align}
Then, for  all $k\ge0$ and all $j=1,\ldots,\ell,$ the requested optimal  control is given by
\begin{equation}
     u_{{\rm req},k|j} = K_{k|j} x_{\rmm,k|j},
     \label{eq:tuKxm}
\end{equation}
where
\begin{equation}
    K_{k|j} \isdef -(R_{2,k|j} +   B_{\rmm,k}^\rmT P_{k|j+1}   B_{\rmm,k})^{-1}  B_{\rmm,k}^\rmT P_{k|j+1}  A_{\rmm,k}.
    \label{eq:Kkjj}
\end{equation}
For all $k\ge0,$ define $K_{k+1}\isdef K_{k|1}$.
Then, for all $k\ge0,$  it follows from \eqref{eq:xmtuk1} and \eqref{eq:tuKxm} that 
\begin{align}
  u_{{\rm req},k+1}  &=  K_{k+1} x_{\rmm,k+1}, 
  \label{eq:ukxkkx}
\end{align}
which, combined with \eqref{eq:Kkjj}, implies that
\small
\begin{equation}
u_{{\rm req},k+1}=  -(R_{2,k|1} +   B_{\rmm,k}^\rmT P_{k|2}  B_{\rmm,k})^{-1}  B_{\rmm,k}^\rmT P_{k|2}   A_{\rmm,k} x_{\rmm,k+1}.
\end{equation}
\normalsize
%
For all $k\ge 0$, the discrete-time control that is implemented at step $k+1$ is given by
\begin{equation}
    u_{k+1}  =\sigma( u_{{\rm req},k+1}),
    \label{eq:uk1h1}
\end{equation}
where $\sigma\colon \BBR^{m}\to\BBR^m$ is given by \eqref{eq:Hsat1}.

Note that the initial control $u_0\in\BBR^m$ is not computed and must be specified. 
In this work, $u_0 = 0.$
In addition, note that, for all $k\ge0,$ the requested control $u_{{\rm req},k+1}$ is computed during the interval $[kT_\rms,(k+1)T_\rms)$ using the measurement $y_k$ and the implemented control $u_k.$
Furthermore, note that, for all $k\ge0$ and all $j = 2,\ldots,\ell,$ $K_{k|j}$, $x_{\rmm,k|j},$ and $u_{{\rm req},k|j}$ need not be computed, in accordance with receding-horizon control. 
Finally, for all examples in this paper, we choose $R_{1,k|j}$, $R_{2,k|j}$, $E_{1,k|j}$ and $P_{k|\ell+1}$  to be independent of $k$ and $j,$ and we thus write $R_1$, $R_2$, $E_1$, and $P_{\ell+1},$ respectively.


\section{Description of the Rijke-tube setup}\label{sec:Rijke_Exp}

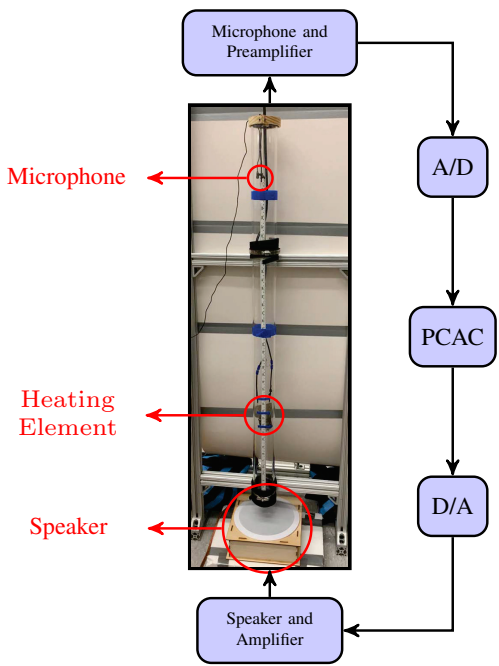
\begin{figure}[h]
\centering
    \resizebox{0.8\columnwidth}{!}{%
    \begin{tikzpicture}[>={stealth'}, line width = 0.25mm]
    	\node [smallblock, inner sep=0.0em, fill=none, line width = 0.75mm] (system) {\centering \includegraphics[width=4.5em]{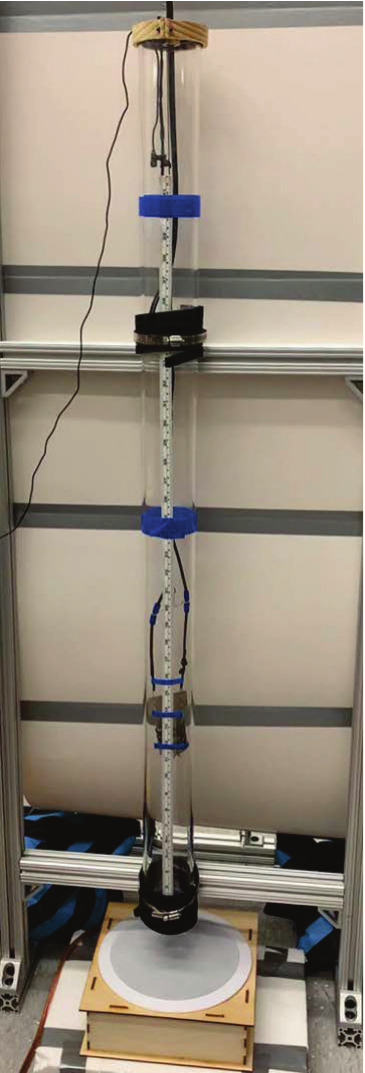}};
        \node [sum, draw = red, fill = none, inner sep = 0.25em,above right = 4.3em and -0.55em of system.center](sum1){};
        \node [sum, draw = red, fill = none, inner sep = 0.39em,below right = 1.85em and -0.58em of system.center](sum2){};
        \node [sum, draw = red, fill = none, inner sep = 0.9em,below right = 4.6em and -1em of system.center](sum3){};
        \draw[->, red] (sum1.west) -- node [xshift = -3.75em]{\scriptsize Microphone} ([xshift = -3.65em]sum1.east);
        \draw[->, red] (sum2.west) -- node [xshift = -3.65em]{\scriptsize $\begin{array}{c} {\rm Heating} \\ {\rm Element} \end{array}$} ([xshift = -4em]sum2.east);
        \draw[->, red] (sum3.west) -- node [xshift = -3.35em]{\scriptsize Speaker} ([xshift = -4.75em]sum3.east);
    	\node [smallblock, rounded corners, minimum height = 0.6cm , minimum width = 0.7cm, above = 7.5em of system.center](micpreamp){\tiny$\begin{array}{c} \mbox{Microphone and} \\ \mbox{Preamplifier} \end{array}$};
    	\node [smallblock, rounded corners, minimum height = 0.6cm , minimum width = 0.7cm, below = 7.5em of system.center](spkamp){\tiny$\begin{array}{c} \mbox{Speaker and} \\ \mbox{Amplifier} \end{array}$};
    	%
    	%
    	\node [smallblock, rounded corners, minimum height = 0.6cm , minimum width = 0.7cm, right = 4em of system.center] (controller){\scriptsize PCAC};
    	\node [smallblock, rounded corners, minimum height = 0.6cm , minimum width = 0.7cm, above = 4em of controller.center] (preamp){\scriptsize A/D};
    	\node [smallblock, rounded corners, minimum height = 0.6cm , minimum width = 0.7cm, below = 4em of controller.center] (lsAmp){\scriptsize D/A};
    	\draw[->](preamp.south)--(controller.north);
    	\draw[->](controller.south)--(lsAmp.north);
    	\draw[->](system.north)--(micpreamp.south);
    	\draw[->](micpreamp.east)-|(preamp.north);
    	\draw[->](lsAmp.south)|-(spkamp.east);
    	\draw[->](spkamp.north)--(system.south);
	\end{tikzpicture}
    }
    \caption{Physical closed-loop Rijke-tube setup.  The heating element can be raised or lowered by a DC motor (not shown) to vary the dynamics of the system.}
    \label{fig:Rijke_tube_Exp}
    \vspace{-1.5em}
\end{figure}

The experimental Rijke-tube setup considered in this work is shown in Figure \ref{fig:Rijke_tube_Exp}, where a heating element is placed inside a vertical Pyrex tube whose length is 1.2 m and whose
inner cross-sectional area
is $4.6 \cdot 10^{-3}$ $\rmm^2,$ similarly to the setup in \cite{epperlein2015}.
The heating element is a coil made from 22-gauge Kanthal wire with a  resistance of 22 ohms and is placed $x_{\rm us}$ m above the bottom of the tube.
The coil is attached by a Kevlar rope to a DC motor, which is used to reposition the coil and modulate $x_{\rm us}.$
A Variac is used as a power supply to modulate the root-mean-square (RMS) voltage $V_{\rm RMS}$ V supplied to the coil.
A microphone is placed at the top of the tube and connected to a preamplifier to measure the resulting acoustic pressure $\tilde{p}_{\rm mic}.$
The microphone was calibrated using a sound pressure level meter to convert voltage measurements to pascals (Pa).
%
%
A speaker is placed at the bottom of the tube and connected to an amplifier so that the predictive controller can modulate the speaker voltage $\tilde{p}_{\rm spk}$ V.
Rijke-tube experiments can be operated in open-loop and closed-loop mode; in open-loop mode, $\tilde{p}_{\rm spk} \equiv 0,$ while, in closed-loop mode, $\tilde{p}_{\rm spk}$ is given by the output of the control algorithm.

\begin{figure}[h!]
    \centering
    \vspace{0.5em}
    \resizebox{\columnwidth}{!}{%
    \includegraphics{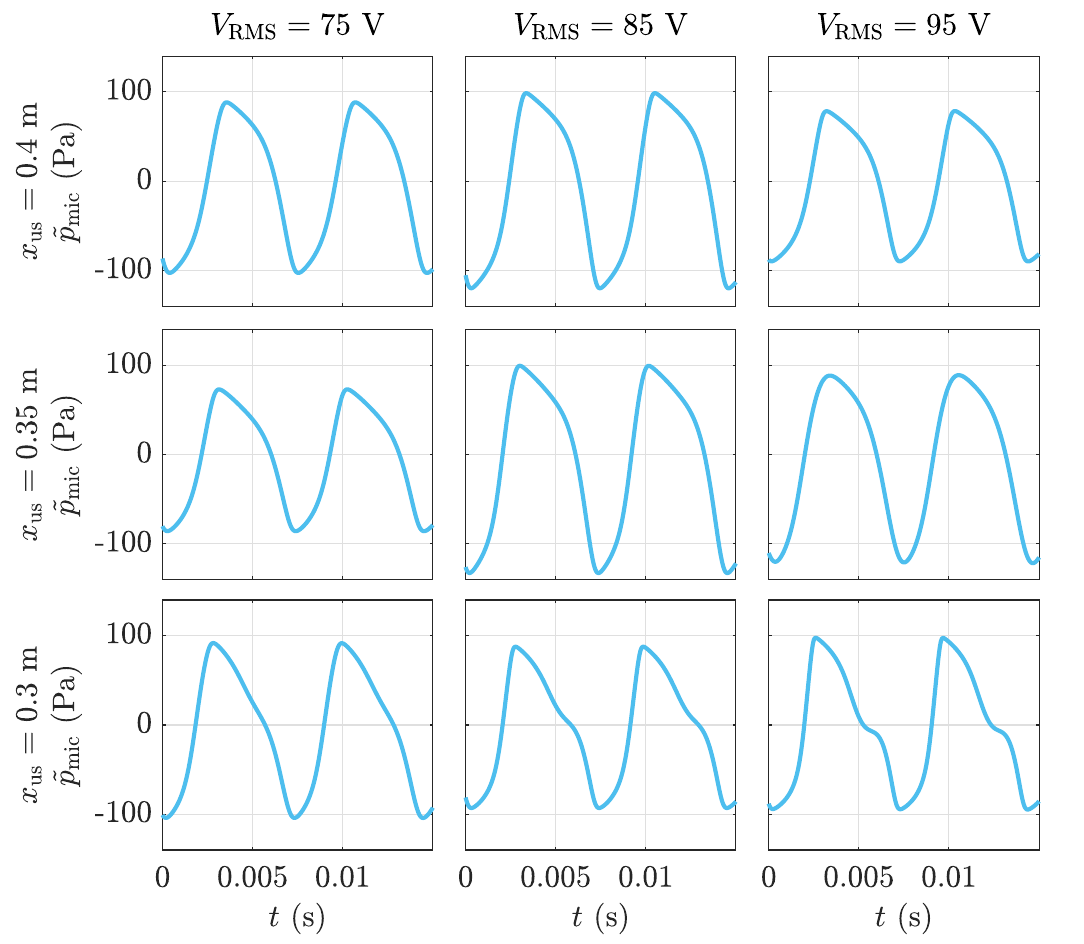}
    }
    \caption{Pressure measurements from the open-loop experimental Rijke-tube setup obtained at the coil positions $x_{\rm us} \in \{0.3, 0.35, 0.4\}$ m and the AC voltage levels $V_{\rm RMS} \in \{75, 85, 95\}$ V, where $x_{\rm us}$ is the distance of the coil from the bottom of the tube, and $V_{\rm RMS}$ is the root-mean-square (RMS) voltage provided by the Variac.
    }
    \label{fig:rijke_tube_exp_OL_time}
    \vspace{-1em}
\end{figure}

\begin{figure}[h!]
    \centering
    \resizebox{\columnwidth}{!}{%
    \includegraphics{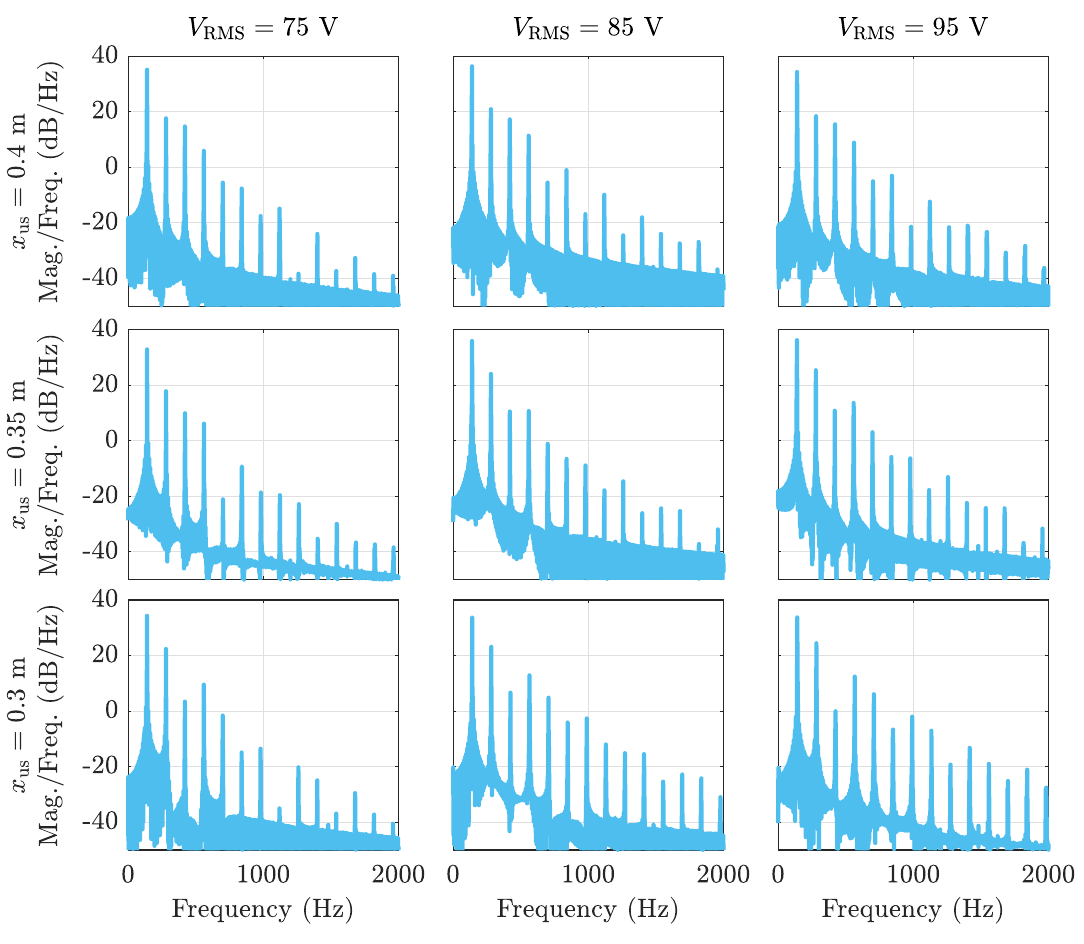}
    }
    \caption{Amplitude spectra of the pressure measurements from the open-loop experiments at each setting considered in Figure \ref{fig:rijke_tube_exp_OL_time}.
    }
    \label{fig:rijke_tube_exp_OL_freq}
    \vspace{-1em}
\end{figure}

Pressure oscillations are created within the experimental Rijke-tube setup by supplying voltage to the heating element, as noted by Rijke in \cite{rijke1859} and subsequently elucidated by Rayleigh \cite{rayleigh1878,sarpotdar2003}.
As explained in \cite{sarpotdar2003,raun1993,manoj2022} \cite[pp. 232-234]{rayleigh1896}, pressure oscillations are created and become self-excited if and only if
the heating element is placed in the lower half of the tube and 
sufficient power is provided to the heating element to overcome the acoustic damping.
Furthermore, pressure oscillations are more easily created when the heat source is placed at one quarter of the length of the tube from its bottom and become harder to create as the heat source is moved from this position \cite{sarpotdar2003}.
The chosen experimental Rijke-tube setup exhibits thermoacoustic oscillations in open-loop mode, whose characteristics depend on the vertical position of the heating element ($x_{\rm us}$) and the voltage provided to the heating element ($V_{\rm RMS}$), as shown in Figures \ref{fig:rijke_tube_exp_OL_time} and \ref{fig:rijke_tube_exp_OL_freq}.

\begin{figure}[h!]
    \centering
    \vspace{0.5em}
    \resizebox{\columnwidth}{!}{%
    \includegraphics{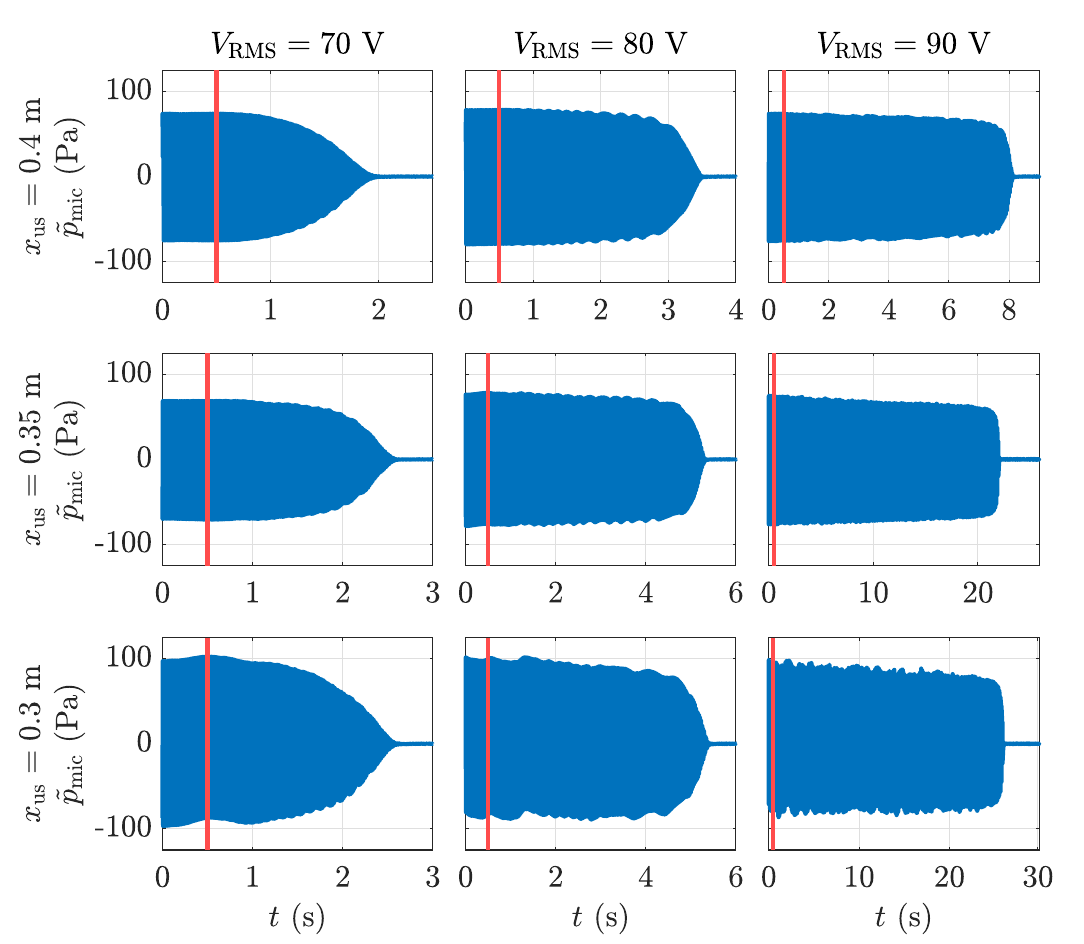}
    }
    \caption{Pressure measurements $\tilde{p}_{\rm mic}$ from the closed-loop experiments using Retrospective Cost Adaptive Control (RCAC) from \cite{RijkeTCST} are shown for $x_{\rm us} \in \{0.3, 0.35, 0.4\}$ m and $V_{\rm RMS} \in \{70, 80, 90\}$ V.
    Each experiment transitions from open-loop mode to closed-loop mode at the time indicated by the vertical red line.
    The same RCAC hyperparameters are used in all tests.
    }
    \label{fig:rijke_tube_exp_RCAC}
    \vspace{-1em}
\end{figure}

In \cite{RijkeTCST}, Retrospective Cost Adaptive Control (RCAC) was used to suppress Rijke-tube thermoacoustic oscillations under various system parameters, as shown in Figure \ref{fig:rijke_tube_exp_RCAC}.
In this work, PCAC is used in Section \ref{sec:Rijke_Exp_Res} to suppress the oscillations for all the cases shown in Figures \ref{fig:rijke_tube_exp_OL_time} and \ref{fig:rijke_tube_exp_OL_freq}, and show that PCAC suppresses the Rijke-tube oscillations faster than RCAC in \cite{RijkeTCST}.

\section{Physical closed-loop experiments using Rijke-tube setup}\label{sec:Rijke_Exp_Res}

In this section, PCAC is implemented in the Rijke-tube setup with a sampling time of $T_\rms = 0.001$ s/step under various system parameters for closed-loop experiments as explained in Section \ref{sec:Rijke_Exp}.
For the Rijke-tube closed-loop experiments, $u \equiv \tilde{p}_{\rm spk},$ and  $y \equiv \tilde{p}_{\rm mic},$
which implies that the Rijke-tube setup is SISO and thus $m = p = 1.$
Furthermore, the initial set of hyperparameters for RLS and BPRE are chosen by following the hyperparameter selection procedure introduced in \cite{RijkeTCST} and slightly modified during experiments to improve suppression performance.
Thus, the hyperparameters for RLS are given by
\begin{gather*}
  \hat{n} = 10,\quad \theta_0 = 10^{-10}\, \mathds{1}_{2 \hat{n}\times 1},\quad \Psi_0 = 10^{-4} I_{2 \hat{n}}, \\
  \tau_\rmn = 40,\quad \tau_\rmd = 200,\quad \eta = 0.1,\quad \alpha = 0.001,
\end{gather*}
and the hyperparameters for BPRE are given by
\begin{gather*}
    \ell = 20, \ P_{\ell+1}= \diag(1, 0_{1\times \hat{n}-1}), \
    R_1 =  \diag(1,0_{1\times \hat{n}-1}), \\
    R_2= 10^{-2} , \ u_{\max} =-u_{\min}=8.
\end{gather*}

We consider experimental scenarios where the coil position and supplied voltage are kept constant.
In total, 9 combinations are considered, such that $x_{\rm us} \in \{0.3, 0.35, 0.4\}$ m and $V_{\rm RMS} \in \{75, 85, 95\}$ V, which are the cases shown in Figures \ref{fig:rijke_tube_exp_OL_time} and \ref{fig:rijke_tube_exp_OL_freq}.
Through testing, it is determined that the oscillations are more difficult to suppress as $x_{\rm us}$ moves closer to 0.3 m (a quarter of the tube length from its bottom, as mentioned in Section \ref{sec:Rijke_Exp}) and $V_{\rm RMS}$ increases.
The experiments begin in open-loop mode to allow the thermoacoustic oscillations to fully develop.
Then, the experiments transition to closed-loop mode, in which PCAC starts modulating the system

\begin{figure}[h!]
    \centering
    \vspace{0.5em}
    \resizebox{\columnwidth}{!}{%
    \includegraphics{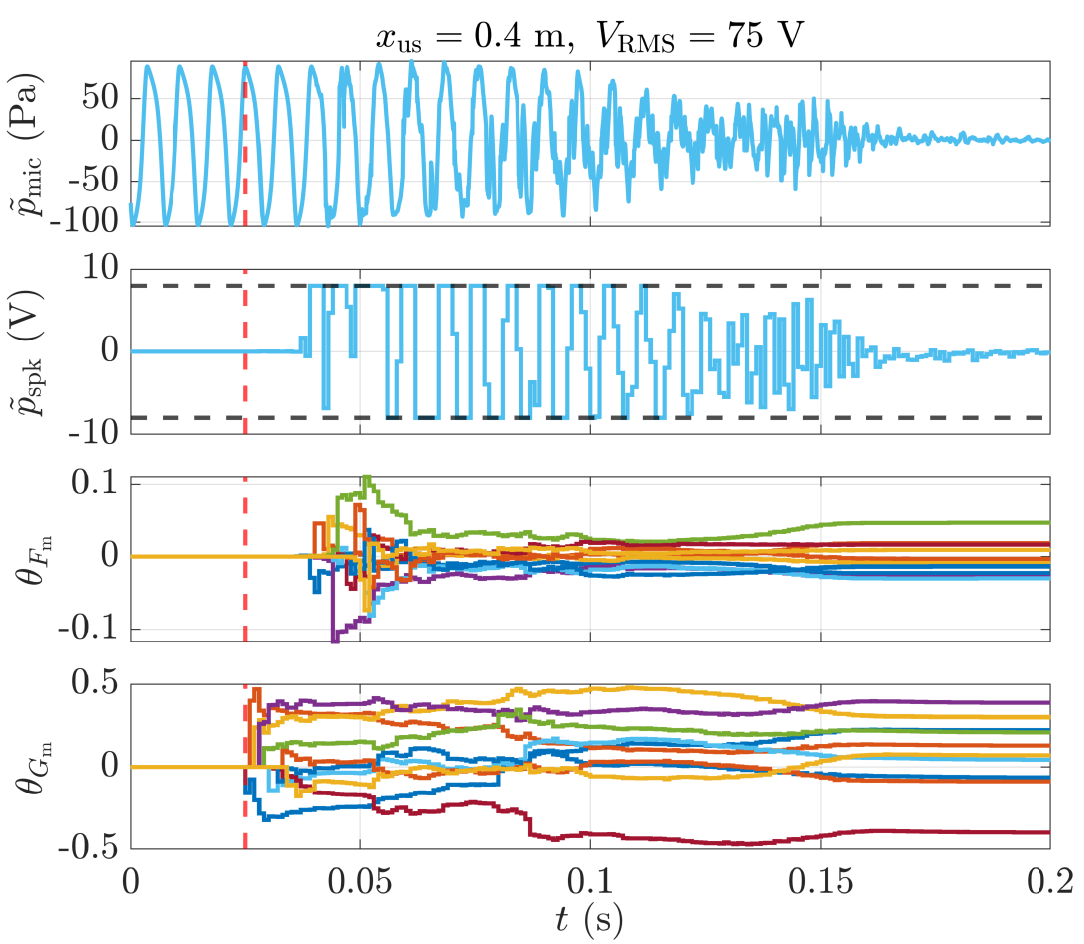}
    }
    \caption{Pressure measurements $\tilde{p}_{\rm mic},$ requested speaker voltage  $\tilde{p}_{\rm spk},$ and estimated model coefficients $\theta_{F_\rmm}$ and $\theta_{G_\rmm}$ from the closed-loop experiments using the predictive controller for $x_{\rm us} = 0.4$ m and $V_{\rm RMS} = 75$ V, for $t \in [0, 0.2]$ s.
    The experiment transitions from open-loop mode to closed-loop mode at the time indicated by the vertical, dashed red line.
    The horizontal, dashed black lines in the $\tilde{p}_{\rm spk}$ versus $t$ plot correspond to the values of $u_{\rm min}$ and $u_{\rm max}.$
    }
    \label{fig:rijke_tube_exp_CL}
    \vspace{-1em}
\end{figure}

\begin{figure}[h!]
    \centering
    \resizebox{\columnwidth}{!}{%
    \includegraphics{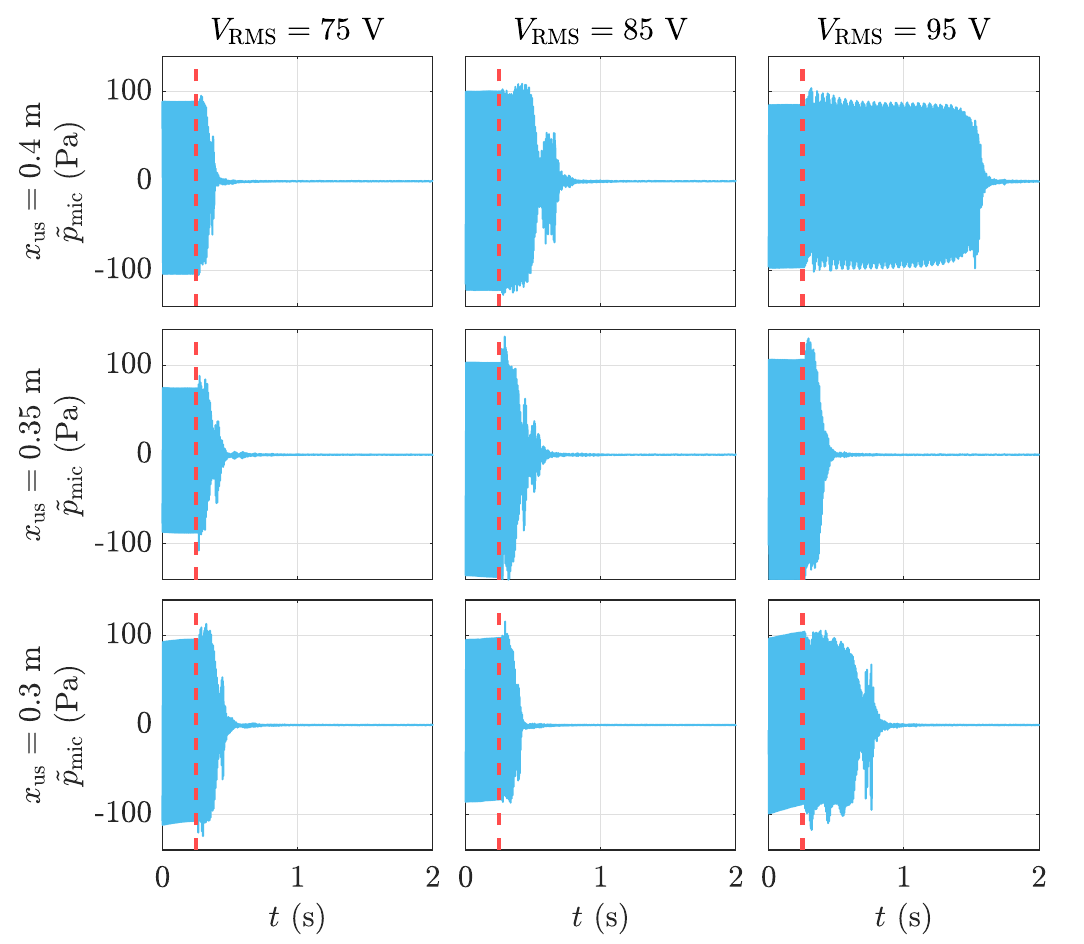}
    }
    \caption{Pressure measurements $\tilde{p}_{\rm mic}$  from the closed-loop experiments using the predictive controller for $x_{\rm us} \in \{0.3, 0.35, 0.4\}$ m and $V_{\rm RMS} \in \{75, 85, 95\}$ V, for $t \in [0, 2]$ s.
    The experiment transitions from open-loop mode to closed-loop mode at the time indicated by the vertical, dashed red line.
    The same PCAC hyperparameters are used in all tests.
    }
    \label{fig:rijke_tube_exp_CL_time}
    \vspace{-1em}
\end{figure}

\begin{figure}[h!]
    \centering
    \vspace{0.5em}
    \resizebox{\columnwidth}{!}{%
    \includegraphics{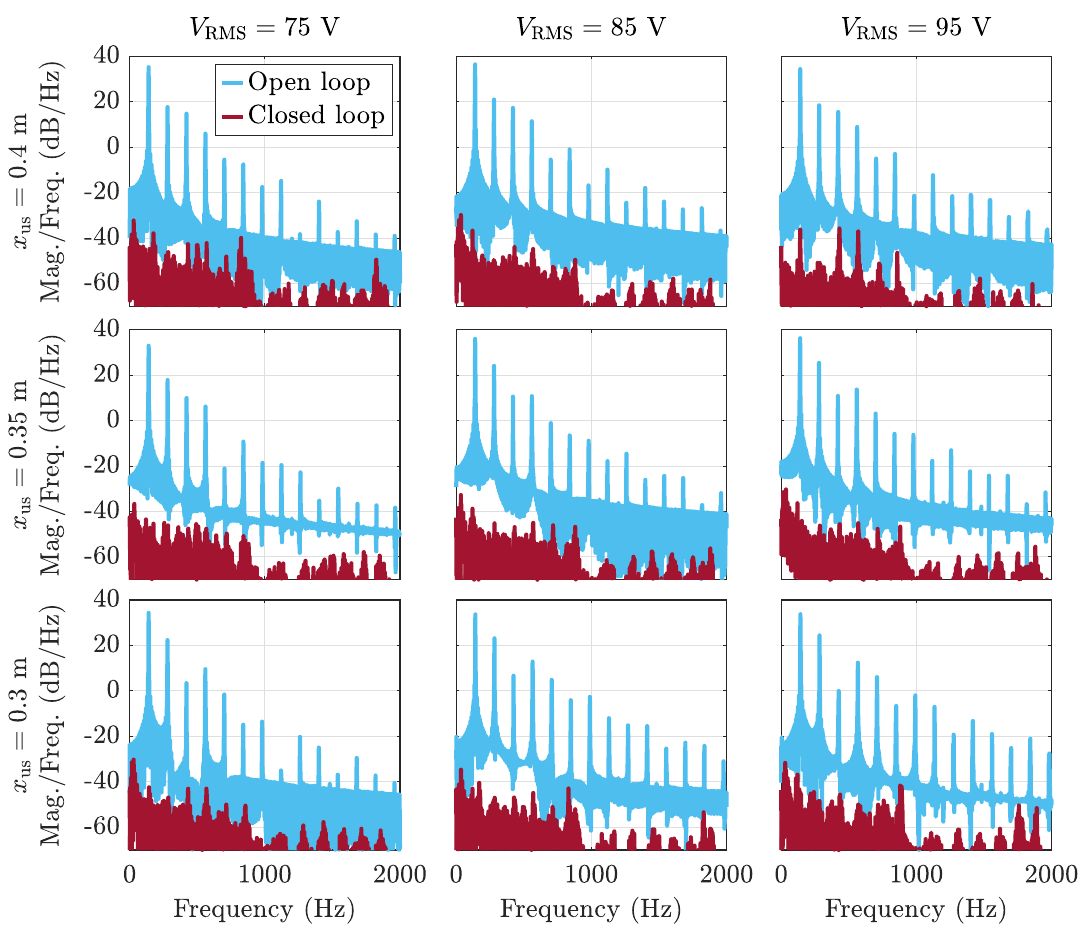}
    }
    \caption{Amplitude spectra of the experimental Rijke-tube setup.
    The amplitude spectra of the pressure measurements obtained from the open-loop experiments and the closed-loop experiments using the predictive controller are shown for $x_{\rm us} \in \{0.3, 0.35, 0.4\}$ m and $V_{\rm RMS} \in \{75, 85, 95\}$ V.
    The same PCAC hyperparameters are used in all tests.
    }
    \label{fig:rijke_tube_exp_CL_freq}
    \vspace{-1em}
\end{figure}

The results of the physical closed-loop experiment in the case where $x_{\rm us} = 0.4$ m and $V_{\rm RMS} = 75$ V are shown in Figure \ref{fig:rijke_tube_exp_CL}, which shows PCAC suppressing the thermoacoustic oscillations in the Rijke tube in less than 0.2 s while respecting the constraints imposed on $\tilde{p}_{\rm spk}$ by $u_{\rm min}$ and $u_{\rm max}.$
Furthermore, the results of the physical closed-loop experiments for $x_{\rm us} \in \{0.3, 0.35, 0.4\}$ m and $V_{\rm RMS} \in \{75, 85, 95\}$ V are shown in Figures \ref{fig:rijke_tube_exp_CL_time} and \ref{fig:rijke_tube_exp_CL_freq},  which show PCAC suppressing the thermoacoustic oscillations in the Rijke tube in all cases in less than 1.5 s.
It can be seen from Figures \ref{fig:rijke_tube_exp_RCAC} and \ref{fig:rijke_tube_exp_CL_time} that suppression is achieved much faster using PCAC than using RCAC in \cite{RijkeTCST} in all cases.

\section{Conclusions}\label{sec:conclusion}

This paper introduced  predictive cost adaptive control for discrete-time, output-feedback control of a continuous-time system.
Then, the Rijke-tube setup was presented and its open-loop response under various system parameters was shown.
Finally, the predictive controller was used to suppress the oscillatory response of the Rijke-tube setup under various system parameters.
The results show that PCAC suppresses the Rijke-tube oscillations faster than RCAC in \cite{RijkeTCST}.
Future work will aim to develop a theoretical framework for guaranteeing stability and convergence under minimal modeling information for self-excited systems.

\section*{Acknowledgments}

This research was supported by ONR under grant N00014-18-1-2211 and AFOSR under grant FA9550-20-1-0028.
The authors thank John Spencer for assistance with the simulations and experiments.

\bibliographystyle{IEEEtran}
\bibliography{IEEEabrv,bib_paper.bib}

\begin{thebibliography}{10}
\providecommand{\url}[1]{#1}
\csname url@samestyle\endcsname
\providecommand{\newblock}{\relax}
\providecommand{\bibinfo}[2]{#2}
\providecommand{\BIBentrySTDinterwordspacing}{\spaceskip=0pt\relax}
\providecommand{\BIBentryALTinterwordstretchfactor}{4}
\providecommand{\BIBentryALTinterwordspacing}{\spaceskip=\fontdimen2\font plus
\BIBentryALTinterwordstretchfactor\fontdimen3\font minus \fontdimen4\font\relax}
\providecommand{\BIBforeignlanguage}[2]{{%
\expandafter\ifx\csname l@#1\endcsname\relax
\typeout{** WARNING: IEEEtran.bst: No hyphenation pattern has been}%
\typeout{** loaded for the language `#1'. Using the pattern for}%
\typeout{** the default language instead.}%
\else
\language=\csname l@#1\endcsname
\fi
#2}}
\providecommand{\BIBdecl}{\relax}
\BIBdecl

\bibitem{rijke1859}
P.~L. Rijke, ``{LXXI}. {N}otice of a new method of causing a vibration of the air contained in a tube open at both ends,'' \emph{Lond. Edinb. Dubl. Phil. Mag}, vol.~17, no. 116, pp. 419--422, 1859.

\bibitem{rayleigh1878}
J.~W.~S. Rayleigh, ``The explanation of certain acoustical phenomena,'' \emph{Nature}, vol.~18, no. 455, pp. 319--321, 1878.

\bibitem{heckl1990}
M.~A. Heckl, ``Non-linear acoustic effects in the {R}ijke tube,'' \emph{Acta Acustica}, vol.~72, no.~1, pp. 63--71, 1990.

\bibitem{bittanti2002}
S.~Bittanti, A.~De~Marco, G.~Poncia, and W.~Prandoni, ``Identification of a model for thermoacoustic instabilities in a {R}ijke tube,'' \emph{IEEE Trans. Contr. Sys. Tech.}, vol.~10, no.~4, pp. 490--502, 2002.

\bibitem{epperlein2015}
J.~P. Epperlein, B.~Bamieh, and K.~J. Astrom, ``Thermoacoustics and the {R}ijke tube: {E}xperiments, identification, and modeling,'' \emph{IEEE Contr. Sys. Mag.}, vol.~35, no.~2, pp. 57--77, 2015.

\bibitem{rubio2015}
J.~Rubio-Hervas, M.~Reyhanoglu, and W.~MacKunis, ``Observer-based sliding mode control of {R}ijke-type combustion instability,'' \emph{J. Low Freq. Noise, Vibr. Active Contr.}, vol.~34, no.~2, pp. 201--217, 2015.

\bibitem{deandrade2020}
G.~A. de~Andrade, R.~Vazquez, and D.~J. Pagano, ``Backstepping-based estimation of thermoacoustic oscillations in a {R}ijke tube with experimental validation,'' \emph{IEEE Trans. Autom. Contr.}, vol.~65, no.~12, pp. 5336--5343, 2020.

\bibitem{heckl1988}
M.~A. Heckl, ``Active control of the noise from a {R}ijke tube,'' \emph{J. Sound Vib.}, vol. 124, no.~1, pp. 117--133, 1988.

\bibitem{annaswamy2000}
A.~M. Annaswamy, M.~Fleifil, J.~W. Rumsey, R.~Prasanth, J.-P. Hathout, and A.~F. Ghoniem, ``Thermoacoustic instability: Model-based optimal control designs and experimental validation,'' \emph{IEEE Trans. Contr. Sys. Tech.}, vol.~8, no.~6, pp. 905--918, 2000.

\bibitem{illingworth2010}
S.~J. Illingworth and A.~S. Morgans, ``Advances in feedback control of the {R}ijke tube thermoacoustic instability,'' \emph{Int. J. Flow Contr.}, vol.~2, no.~4, 2010.

\bibitem{zalluhoglu2016}
U.~Zalluhoglu, A.~S. Kammer, and N.~Olgac, ``Delayed feedback control laws for {R}ijke tube thermoacoustic instability, synthesis, and experimental validation,'' \emph{IEEE Trans. Contr. Sys. Tech.}, vol.~24, no.~5, pp. 1861--1868, 2016.

\bibitem{deandrade2017}
G.~A. de~Andrade, R.~Vazquez, and D.~J. Pagano, ``Boundary control of a rijke tube using irrational transfer functions with experimental validation,'' in \emph{Proc. IFAC World Congress}, 2017, pp. 4528--4533.

\bibitem{juanACCrijke}
J.~Paredes, S.~A.~U. Islam, and D.~S. Bernstein, ``Adaptive stabilization of thermoacoustic oscillations in a {R}ijke tube,'' in \emph{Proc. Amer. Contr. Conf.}, 2022, pp. 28--33.

\bibitem{RijkeTCST}
J.~Paredes and D.~S. Bernstein, ``{Experimental Implementation of Retrospective Cost Adaptive Control for Suppressing Thermoacoustic Oscillations in a Rijke Tube},'' \emph{IEEE Trans. Contr. Sys. Tech.}, 2023, dOI: 10.1109/TCST.2023.3262223.

\bibitem{lee2011}
J.~H. Lee, ``{Model predictive control: Review of the three decades of development},'' \emph{Int. J. Contr. Autom. Sys.}, vol.~9, pp. 415--424, 2011.

\bibitem{darby2012}
M.~L. Darby and M.~Nikolaou, ``{MPC: Current practice and challenges},'' \emph{Contr. Eng. Pract.}, vol.~20, no.~4, pp. 328--342, 2012.

\bibitem{afram2014}
A.~Afram and F.~Janabi-Sharifi, ``{Theory and applications of HVAC control systems--A review of model predictive control (MPC)},'' \emph{Build. Environ.}, vol.~72, pp. 343--355, 2014.

\bibitem{incremona2017}
G.~P. Incremona, A.~Ferrara, and L.~Magni, ``{MPC for robot manipulators with integral sliding modes generation},'' \emph{Trans. Mechatronics}, vol.~22, no.~3, pp. 1299--1307, 2017.

\bibitem{torrente2021}
G.~Torrente, E.~Kaufmann, P.~F{\"o}hn, and D.~Scaramuzza, ``{Data-driven MPC for quadrotors},'' \emph{Rob. Autom. Lett.}, vol.~6, no.~2, pp. 3769--3776, 2021.

\bibitem{aliramezani2022}
M.~Aliramezani, C.~R. Koch, and M.~Shahbakhti, ``Modeling, diagnostics, optimization, and control of internal combustion engines via modern machine learning techniques: A review and future directions,'' \emph{Prog. Energ. Comb. Sci.}, vol.~88, p. 100967, 2022.

\bibitem{jarmolowitz2012conf}
F.~Jarmolowitz, C.~Gro{\ss}-Weege, T.~Lammersen, S.~Shariati, and D.~Abel, ``{Modelling and robust Model Predictive Control of an unstable thermoacoustic system with constraints},'' in \emph{Proc. Amer. Contr. Conf.}\hskip 1em plus 0.5em minus 0.4em\relax IEEE, 2012, pp. 6588--6595.

\bibitem{jarmolowitz2012journal}
F.~Jarmolowitz, C.~Gro{\ss}-Weege, T.~Lammersen, D.~Abel \emph{et~al.}, ``{Robust output model predictive control of an unstable Rijke tube},'' \emph{J. Comb.}, vol. 2012, 2012.

\bibitem{shariati2014}
S.~Shariati, A.~A. da~Franca, B.~Oezer, R.~Noske, D.~Abel, and A.~Brockhinke, ``{Modeling and model predictive control of combustion instabilities in a multi-section combustion chamber using two-port elements},'' in \emph{Proc. Conf. Contr. Appl.}\hskip 1em plus 0.5em minus 0.4em\relax IEEE, 2014, pp. 2108--2113.

\bibitem{islam2019recursive}
S.~A.~U. Islam and D.~S. Bernstein, ``Recursive least squares for real-time implementation,'' \emph{IEEE Contr. Syst. Mag.}, vol.~39, no.~3, pp. 82--85, 2019.

\bibitem{mohseni2022recursive}
N.~Mohseni and D.~S. Bernstein, ``Recursive least squares with variable-rate forgetting based on the {F}-test,'' in \emph{Proc. Amer. Contr. Conf.}, 2022, pp. 3937--3942.

\bibitem{islamPCAC}
T.~W. Nguyen, S.~A.~U. Islam, D.~S. Bernstein, and I.~V. Kolmanovsky, ``{Predictive Cost Adaptive Control: A Numerical Investigation of Persistency, Consistency, and Exigency},'' \emph{IEEE Contr. Sys. Mag.}, vol.~41, pp. 64--96, December 2021.

\bibitem{kwon2006receding}
W.~Kwon and S.~Han, \emph{Receding Horizon Control: Model Predictive Control for State Models}.\hskip 1em plus 0.5em minus 0.4em\relax Springer, 2006.

\bibitem{kwonpearson}
W.~H. Kwon and A.~E. Pearson, ``On feedback stabilization of time-varying discrete linear systems,'' \emph{IEEE Trans. Autom. Contr.}, vol. AC-23, no.~3, pp. 479--481, 1978.

\bibitem{polderman1989state}
J.~W. Polderman, ``A state space approach to the problem of adaptive pole assignment,'' \emph{Mathematics of Control, Signals and Systems}, vol.~2, no.~1, pp. 71--94, 1989.

\bibitem{sarpotdar2003}
S.~M. Sarpotdar, N.~Ananthkrishnan, and S.~Sharma, ``The {R}ijke tube--{A} thermo-acoustic device,'' \emph{Resonance}, vol.~8, no.~1, pp. 59--71, 2003.

\bibitem{raun1993}
R.~Raun, M.~Beckstead, J.~Finlinson, and K.~Brooks, ``A review of {R}ijke tubes, {R}ijke burners and related devices,'' \emph{Prog. Energ. Comb. Sci.}, vol.~19, no.~4, pp. 313--364, 1993.

\bibitem{manoj2022}
K.~Manoj, S.~A. Pawar, J.~Kurths, and R.~Sujith, ``{R}ijke tube: {A} nonlinear oscillator,'' \emph{Chaos}, vol.~32, no.~7, p. 072101, 2022.

\bibitem{rayleigh1896}
J.~W.~S. Rayleigh, \emph{{T}he {T}heory of {S}ound}.\hskip 1em plus 0.5em minus 0.4em\relax Macmillan \& Company, 1896, vol.~2.

\end{thebibliography}

\end{document}